# ANXIETY AS A FACTOR IN THE DEVELOPMENT OF AUTISTIC SPECTRUM SYMPTOMS: AN EXPERIMENTAL STUDY

**I.L. Kovalenko, N.N. Kudryavtseva**

*Neurogenetics of Social Behavior Sector, Institute of Cytology and Genetics SD RAS, Novosibirsk*

It is well known that psychoemotional disorders may be accompanied by decreased sociability in humans. It has been shown that repeated social defeats in 10 daily agonistic interactions in male mice led to development of the expressed level of anxiety and to reduction of communication estimated in the elevated plus-maze and partition tests, respectively. In the social interactions test, sociable behavior toward unfamiliar partner and exploratory activity were dramatically decreased in defeated male mice. Avoidance of approaching partner was significantly increased. Demonstration of self-grooming behavior was increased in defeated males. Chronic diazepam treatment (0.5 mg/kg, i.p., 2 weeks) led to significant decrease of anxiety level estimated in the elevated plus maze test and to improvement of communication in the partition test. In the social interaction test diazepam completely restored the level of sociability and exploratory activity and reduced avoidance behavior of approaching partner. Diazepam did not affect self-grooming demonstration. It is concluded that main factor which provokes decrease in communication in defeated male mice is the high level of anxiety. Similarity between changes in social behavior in defeated mice and symptoms of autism in humans is discussed. It is supposed that this behavioral approach may be useful for studying the molecular mechanisms of autistic spectrum disorders, developing under chronic negative social experiences.

---
*Correspondence to: Kudryavtseva N.N., av. ac. Lavrentjeva, 10 , Institute of Cytology and Genetics SD RAS, 630090,. Novosibirsk, Russia, e-mail: natnik@.bionet.nsc.ru*

# ВВЕДЕНИЕ

Способность взаимодействовать с себе подобными является необходимой составляющей социального поведения и нормального развития индивида, живущего в сообществе. Выраженные нарушения коммуникативности могут быть следствием аутизма, который относится к болезням развития нервной системы и проявляется, в основном, у детей в раннем возрасте [обзор, 18]. Аутизм характеризуется отклонениями в социализации и общении, а также ограниченным, повторяющимся поведением [4]. Полагают, что в развитии аутизма большую роль играют гены – взаимодействие многих генов, либо редкие мутации, оказывающие сильный эффект в пренатальном или раннем постнатальном периоде [обзоры, 3, 12, 28]. Однако многократно возросшие за последние годы случаи этого тяжелого заболевания невозможно объяснить только генетическими мутациями или наследственностью [обзор, 26]. Значит необходимо принимать во внимание негативное влияние окружающей среды. Кроме того, симптомы аутистического спектра могут появляться при различных заболеваниях [4] и быть следствием нарушений психоэмоциональной сферы или возникать после негативного социального опыта. В этом случае при определенном развитии событий нарушение коммуникативности может сохраняться длительное время, принимая хронический характер.

В настоящее время экспериментальные исследования сосредоточены на поисках адекватных моделей аутизма, для чего изучается социабельность мышей разных линий [обзоры, 11, 13] или же мышей, например, лишенных какого-либо гена (нокауты) [обзоры 13, 25]. Коммуникативность животных оценивается по времени поведенческого реагирования на другую особь в условиях домашней клетки или на нейтральной территории [22, 25]. Если животные одной линии меньше реагировали на тестера, чем особи других линий, то предполагалось, что эта линия может быть использована для изучения механизмов аутизма и поисков способов его лечения. Однако целенаправленного исследования состояний, сопровождающихся появлением симптомов аутистического спектра у животных, возникших под влиянием негативных факторов социальной природы, в настоящее время не проводится в виду отсутствия экспериментальных моделей.

Нами было показано, что хронический социальный стресс, вызванный повторным опытом социальных поражений в ежедневных агрессивных взаимодействиях, приводит к развитию тревожности и снижению коммуникативности у побежденных самцов [6, 23]. Поведенческая реакция на партнера в соседнем отсеке экспериментальной клетки, разделенной прозрачной перегородкой с отверстиями у них была существенно ниже по сравнению с поведением контрольных животных. Побежденные самцы переставали реагировать в этой ситуации, как на знакомого, так и на незнакомого партнера [обзор, 22], возможно, в результате обучения негативному социальному опыту при взаимодействии с агрессивным партнером в условиях общей клетки, в которой проходили агонистические взаимодействия. В этой работе были поставлены другие вопросы: Влияет ли высокий уровень тревожности у мышей на коммуникативность в новых условиях, не несущих угрозу для особи? Если да, то можно ли нарушения в общительности у мышей рассматривать в качестве симптомов аутистического спектра? Кроме того, необходимо было понять, является ли состояние тревоги главным фактором нарушения социальных взаимодействий?

## МАТЕРИАЛЫ И МЕТОДЫ

Животных разводили и содержали в стандартных условиях вивария Института цитологии и генетики СО РАН при световом режиме 12:12 часов (свет:темнота). Стандартный гранулированный корм (НПО «Сельскохозяйственные технологии», Москва) и воду они получали *ad libitum*. После отсаживания от матерей в возрасте 1 мес. мышей содержали в группе по 8-10 особей в клетках 36х23х12 см до периода половой зрелости. Эксперименты проводили на самцах мышей линии C57BL/6J в возрасте 2,5-3 мес и массой тела 27-32 г. Все эксперименты с животными проводили в соответствии с международными правилами (European Communities Council Directive of November 24, 1986 (86/609/EEC).

### *Формирование тревожного состояния у самцов мышей*

Для формирования состояния тревожности у самцов мышей использовали модель сенсорного контакта [21]. Животных попарно помещали в экспериментальные клетки, разделенные пополам прозрачной перегородкой с отверстиями, позволявшей мышам видеть, слышать, воспринимать запахи друг друга (сенсорный контакт), но предотвращавшей физическое взаимодействие. Ежедневно во второй половине дня (14.00-17.00 часов) убирали перегородку, что проводило к межсамцовым конфронтациям. Во время первых 2-3 тестов выявляли победителей (агрессоров), и особей, терпящих поражения (побежденные самцы, жертвы) при взаимодействии с одним и тем же партнером. В дальнейшем после теста побежденного самца пересаживали в новую клетку к незнакомому агрессивному партнеру, сидящему за перегородкой. Если интенсивные атаки со стороны нападающей особи во время агрессивных столкновений длились более 3-х минут, взаимодействие самцов прекращали, вновь устанавливая между ними перегородку. В других ситуациях тест продолжался 10 минут. В результате, после 10-ти дней социальных поражений в агонистических взаимодействиях у побежденных особей развивается высокий уровень тревожности [6].

### *Поведенческие тесты*

**Тест «перегородка»** [обзор, 22] количественно оценивает коммуникативность мышей по поведенческой активности возле прозрачной перфорированной перегородки, разделяющей на две половины общую клетку в реакции на партнера в соседнем отсеке. В данном тесте регистрировали число подходов к перегородке, за которой находился знакомый агрессивный самец, длительность нахождения возле нее в течение 5-минутного наблюдения и среднее время пребывания возле перегородки за один подход. Затем на место знакомого самца (агрессора) подсаживали незнакомого группового самца, и регистрировали те же параметры поведения в течение еще 5 минут. При этом фиксировали периоды, когда экспериментальные мыши находились возле перегородки, принюхиваясь к соседу, касались передними лапами или перемещались вдоль нее, грызли отверстия, демонстрируя желание преодолеть перегородку.

**Тест «приподнятый крестообразный лабиринт»** (далее ПКЛ), поведение в котором чувствительно к действию анксиогенных и анксиолитических препаратов, является одним из наиболее общепринятых тестов для оценки тревожного состояния у грызунов [29]. Лабиринт приподнят над полом на 50 см и состоит из двух открытых и двух закрытых (огороженных с трех сторон) рукавов. Тестирование проводили в течение 5 минут в слабо освещенной комнате. Животное помещали в центр лабиринта, носом в закрытый рукав, и регистрировали время (длительность) нахождения в открытых рукавах, центре и закрытых рукавах лабиринта (данные представлены в процентах от общего времени тестирования); число выходов/входов в открытые рукава, центр и закрытые рукава, выраженное, как и время, в процентном соотношении; общее число входов/выходов в рукава и центр; число переходов из одного закрытого рукава в другой; число заглядываний под лабиринт и выглядываний из закрытого рукава. После каждого животного лабиринт мыли в нескольких водах и высушивали чистыми салфетками.

*Тест «социальные взаимодействия»* был нами разработан для количественной оценки социального поведения животных по отношению к стандартному тестеру - незнакомому партнеру той же линии, содержащемуся до теста в группе. Тест является модификацией теста для крыс, который использовался ранее для выявления анксиолитических свойств препаратов [14]. Исследуемую особь и тестера сажали на нейтральную территорию - в чистую клетку размером 36х23х12 см, помещая их одновременно в противоположные углы клетки. Эта клетка обычно используется для содержания животных и потому не является для них пугающей, в ней не было подстилки, корма и каких-либо других предметов, которые бы отвлекали экспериментальную особь от партнера. При тестировании выделяли следующие формы социального поведения: 1) Избегание стандартного тестера или же замирание при его подходе; 2) общительность - поведение, направленное на приближение к стандартному тестеру - подходы, обнюхивание партнера, следование за ним; 3) исследовательская активность (вставания на задние лапы); 4) аутогруминг, рассматриваемый обычно в качестве показателя смещенной активности в новых условиях. Оценивали число и суммарное время проявления поведенческих актов за время теста. Длительность теста равнялась 10 минутам. После каждого животного клетку мыли в нескольких водах и высушивали чистыми салфетками.

Общим правилом в наших экспериментах является использование периода активации перед тестированием поведения, для чего животных приносили в специальную комнату, заменяли обычную крышку клетки на прозрачное оргстекло, и оставляли их на 5 минут для привыкания к условиям другого освещения и активации. Во время теста осуществлялась видеозапись поведения животных с последующей обработкой видеоматериалов.

Для ответа на вопрос, является ли тревожность фактором нарушения социальных взаимодействий, был использован анксиолитик диазепам с отслеживанием его влияния на особенности коммуникативного поведения на фоне снижения тревожности. Диазепам (в/б, 0.5 мг/кг, Polfa Tarchomin S.A.) начинали вводить мышам после 10 дней социальных взаимодействий. На период введения препаратов агонистические взаимодействия прекращались. Сравнивали три группы животных: побежденные самцы с хроническим введением физиологического раствора и диазепама в течение 14 дней, а также контрольные особи без опыта социальных взаимодействий.

Статистическая обработка данных была проведена с помощью критерия Стьюдента и корреляционного анализа данных. Уровень достоверности считали значимым при $p \leq 0.05$. В экспериментальных группах было 12-13 животных.

## РЕЗУЛЬТАТЫ ИССЛЕДОВАНИЯ

*Тест «Перегородка».* Исследование показало, что у побежденных самцов после 10 дней негативного социального опыта число подходов к перегородке в реакции как на знакомого, так и незнакомого самца не отличалось существенно от контроля ($p > 0.05$). В то же время общее и среднее время пребывания возле перегородки у них были

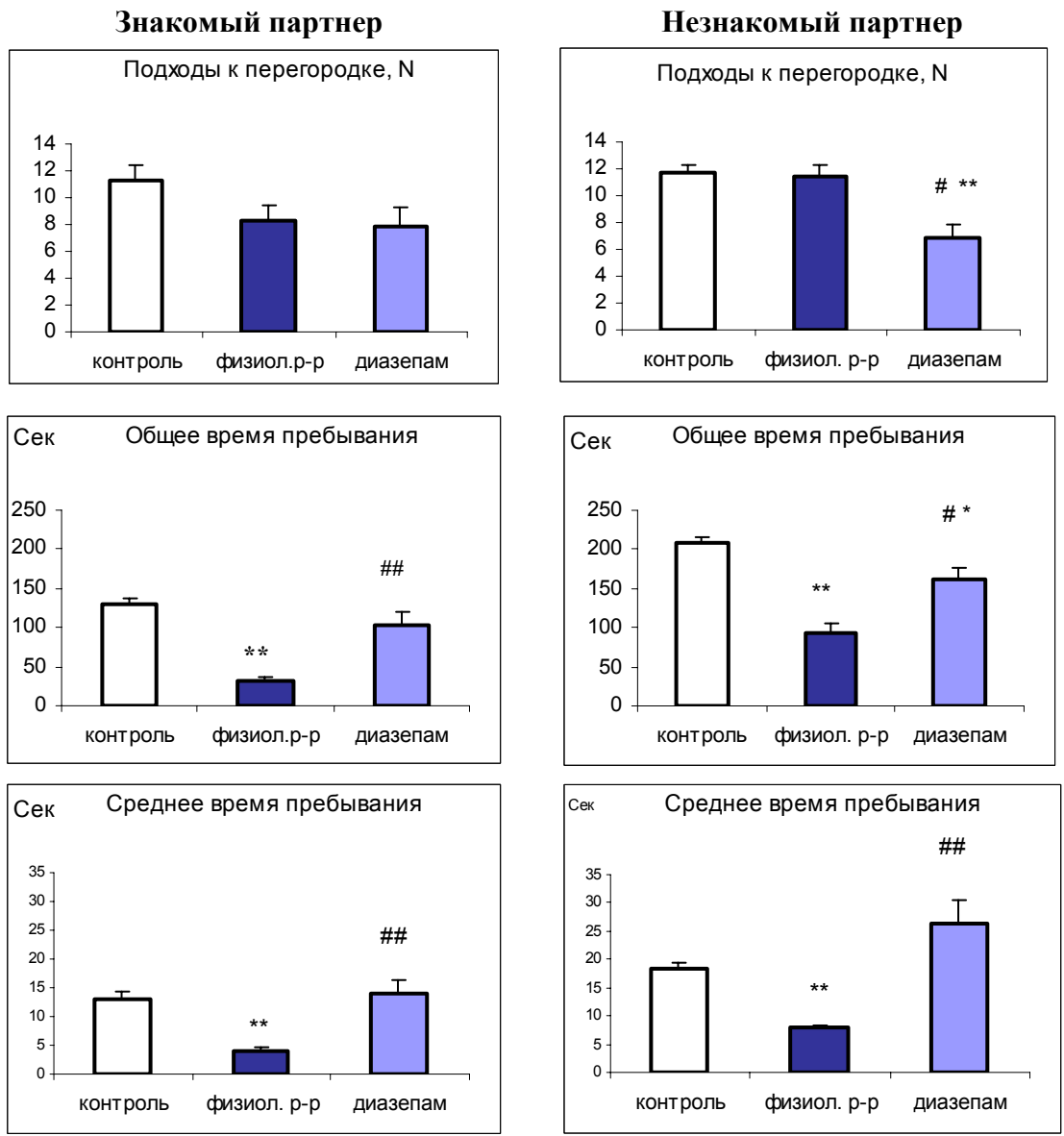

Рис. 1. Поведение самцов мышей экспериментальных групп (контроль, побежденные самцы после хронического введения физиологического раствора и диазепама) в тесте «перегородка» в реакции на знакомого и незнакомого партнера
\* - p < 0.5,  \*\* - p<0,001 – по сравнению с контролем; # - p < 0.01;  ## - p < 0.001  – по сравнению с побежденными самцами, получавшими физиологический раствор

Figure. 1. Behavior of the controls (контроль) and defeated mice after chronic saline (физиол. р-р) and diazepam (диазепам) treatment in the partition test in the reaction to familiar (знакомый) and unfamiliar (незнакомый) partners.
\* - p < 0.5, \*\* - p < 0,001 – vs controls;  # - p < 0.5,  ## - p < 0.001  – vs defeated mice, treated with saline

существенно ниже (для всех показателей *p* < 0.001) *(*рис. 1)**.**  Хроническое введение диазепама существенно снизило число подходов к перегородке в реакции на незнакомого партнера по сравнению с числом подходов у контроля и побежденных самцов, которым вводили физиологический раствор (*p* < 0.001 и *p* < 0.01, соответственно). После введения диазепама общее время нахождения возле перегородки в реакции на знакомого партнера, а также среднее время реакции на знакомого и незнакомого партнера значительно повысились (для всех показателей *p* < 0.001) и не отличались существенно от контроля (*p* > 0.05). После введения диазепама общее время реакции на незнакомого партнера становилось существенно больше у побежденных самцов по сравнению с поведением после введения физиологического раствора (*p* < 0.01). Однако значение этого показателя все же не достигало контрольного уровня (*p* < 0.05).

***Приподнятый крестообразный лабиринт.*** Опыт социальных поражений привел к изменению основных параметров поведения в тесте ПКЛ, оценивающих состояние тревожности. По сравнению с контролем, число выходов и время, проведенное в открытых рукавах стали существенно ниже (для обоих показателей *p* < 0.001), также как число выходов и время пребывания в центре (*p* < .05 и *p* < 0.01, соответственно) (Таблица). Эти показатели для закрытых рукавов у побежденных самцов стали существенно выше (для обоих показателей *p* < 0.001). Число выглядываний из закрытого рукава, число переходов из одного закрытого рукава в другой и число заглядываний под лабиринт у побежденных самцов стали существенно меньше (*p* < 0.01; *p* < 0.001 и *p* < .05, соответственно) по сравнению с контролем.

Таблица   Поведение самцов мышей экспериментальных групп в тесте «приподнятый крестообразный лабиринт» (plus maze test)

| Параметры поведения | Контроль Controls | Побежденные самцы физиол. р-р (saline) | Побежденные самцы диазепам (diazepam) |
|---|---|---|---|
| Открытые рукава, %, N | 10,7±1,5 | 2,1±1,2 *** | 6,1±1,2 * # |
| Открытые рукава, %, сек | 4.9±1.2 | 0.2±0.1 *** | 1.9±0.4 *## |
| Центр, %, N | 45.9±0.7 | 35.2±4.9* | 44.4±1.0 |
| Центр, %, сек | 13.2±1.0 | 6.0±1.7 ** | 10.8±1.6 |
| Закрытые рукава, %, N | 43.4±1.5 | 62.7±4.9 *** | 49.5±1.6 ** # |
| Закрытые рукава, %, сек | 81.8±1.7 | 93.8±1.7 *** | 87.2±1.8 * # |
| Выглядывания, N | 14.2±1.0 | 9.7±1.2 ** | 7.9±1.1 *** |
| Переходы, N | 9.2±0.7 | 2.5±1.1 *** | 4.9±0.9 ** |
| Заглядывания, N | 5.3±1.1 | 2.0±0.6 * | 0.8±0.2 *** |

\* - p < 0.5, *** - p < 0.01, *** - p < 0.001 – по сравнению с контролем (vs controls); # - p < 0.5, ## - p < 0.001; – по сравнению с побежденными самцами, получавшими физиологический раствор (vs saline treated defeated males)

По сравнению с введением физиологического раствора у побежденных животных под влиянием диазепама значительно улучшились: число выходов (p < 0.05) и продолжительность пребывания (*p* < 0.001) в открытых рукавах, число заходов и время пребывания в закрытых рукавах (для обоих показателей *p* < 0.05). Однако по сравнению с контролем у побежденных самцов, которым вводили диазепам, были обнаружены различия по числу заходов и времени пребывания в открытых (для обоих показателей *p* < 0.05) и закрытых (*p* < 0.01 и *p* < 0.05, соответственно) рукавах лабиринта. Число выходов и время пребывания в центре были восстановлены полностью до контрольного уровня – не было обнаружено различий по этим показателям от контроля ($p > 0.05$). По сравнению с контролем сохранялись различия после хронического введения диазепама по числу выглядываний из закрытых рукавов ($p < 0.001$), переходов из одного закрытого рукава в другой ($p < 0.01$) и числу заглядываний под лабиринт ($p < 0.001$).

*Социальные взаимодействия.* Для того, чтобы оценить характер поведения мышей на нейтральной территории были выделены четыре формы поведения, которые, на наш взгляд, проходят на фоне разных мотивационных составляющих – общительность, избегание, вставания на задние лапы и аутогруминг. Прежде всего, необходимо было определить ассоциативность этих форм поведения, то есть насколько они являются взаимосвязанными и/или отражают разные формы поведения животных. Корреляционный анализ не выявил достоверной взаимосвязи между проявлениями таких форм поведения у интактных мышей как: избегание – общительность, избегание - аутогруминг, избегание – вставание на задние лапы, общительность – аутогруминг, общительность – вставание на задние лапы, аутогруминг - вставание на задние лапы (p > 0.05). Достоверные коррелятивные взаимосвязи были обнаружены между числом актов проявления общительности и вставаний на задние лапы (R=0,697; p<0.05), что свидетельствует об общей мотивационной компоненте этих форм поведения - исследовательской активности.

Для того, чтобы определить в какой степени выделенные нами формы коммуникативного поведения отражают уровень тревожности, был проведен корреляционный анализ основных поведенческих параметров в тестах «социальные взаимодействия» и ПКЛ. Оказалось, что у контрольных животных достоверных корреляций обнаружено не было, за исключением одной – между числом демонстрации поведения избегания и временем, проведенным в центре лабиринта (R = 0.638; $p < 0.05$). У животных, потерпевших социальные поражения были обнаружены достоверные корреляции между временем избегания и числом заходов в закрытые рукава (R = 0.805; $p = 0.002$) и в центр (R = - 0.746; $p = 0.005$). Число и время демонстрации общительности позитивно коррелировали с числом выходов в центр (R = 0.617; $p < 0.033$; R = 0.658; $p = 0.020$, соответственно) и отрицательно с числом заходов в закрытые рукава (R = -0.633; $p = 0.027$; R = - 0.595; $p = 0.041$, соответственно). У побежденных самцов, которым вводили диазепам не было обнаружено каких-либо достоверных корреляций между параметрами теста «социальные взаимодействия» и тестом ПКЛ ($p > 0.05$).

Корреляционный анализ основных поведенческих параметров в тестах «социальные взаимодействия» и «перегородка» показал, что у интактных животных достоверно коррелирующими оказались время демонстрации общительности и общее время, проведенное возле перегородки в реакции на незнакомого партнера (R = 0.766, $p$ = 0.010). У побежденных животных были обнаружены достоверные корреляции между числом избеганий и средним временем, проведенным в реакции на знакомого (R = 0.749, $p = 0.005$) и незнакомого (R = 0.579, $p = 0.048$) партнера. У побежденных животных, которым вводили диазепам, были найдены достоверные отрицательные корреляции между общим временем, проведенным возле перегородки в реакции на незнакомого партнера и числом избеганий (R = -0.692, $p = 0.009$), а также временем демонстрации аутогруминга (R = -0.743, $p = 0.004$).

Большую часть времени регистрируемых форм поведения интактные животные демонстрировали вставание на задние лапы и общительность (39% и 42%, соответственно), избегание партнера и аутогруминг у них был выражен в очень незначительной степени (7% и 11%, соответственно). У побежденных самцов, наоборот, в поведении существенно преобладали аутогруминг (26%) и избегание (49%). Время проявления общительности и вставание на задние лапы составляли 9% и 15%, соответственно. Диазепам эти соотношения приближал к норме (вставание на задние лапы – 30%, общительность - 36%, аутогруминг – 20% и избегание - 14%).

Эти соотношения между выделенными формами поведения сохранялись и по абсолютным показателям (Рисунок 2). По сравнению с контролем, у побежденных самцов было существенно больше число актов и время демонстрации поведения избегания (для обоих показателей, $p < 0.001$), а также длительности демонстрации аутогрумингов ($p \leq 0,05$). Показатели общительности - число подходов к партнеру и время пребывания возле него были существенно меньше (для обоих показателей, $p<0.001$), также как и число и длительность вставаний на задние лапы ($p < 0.01$ и

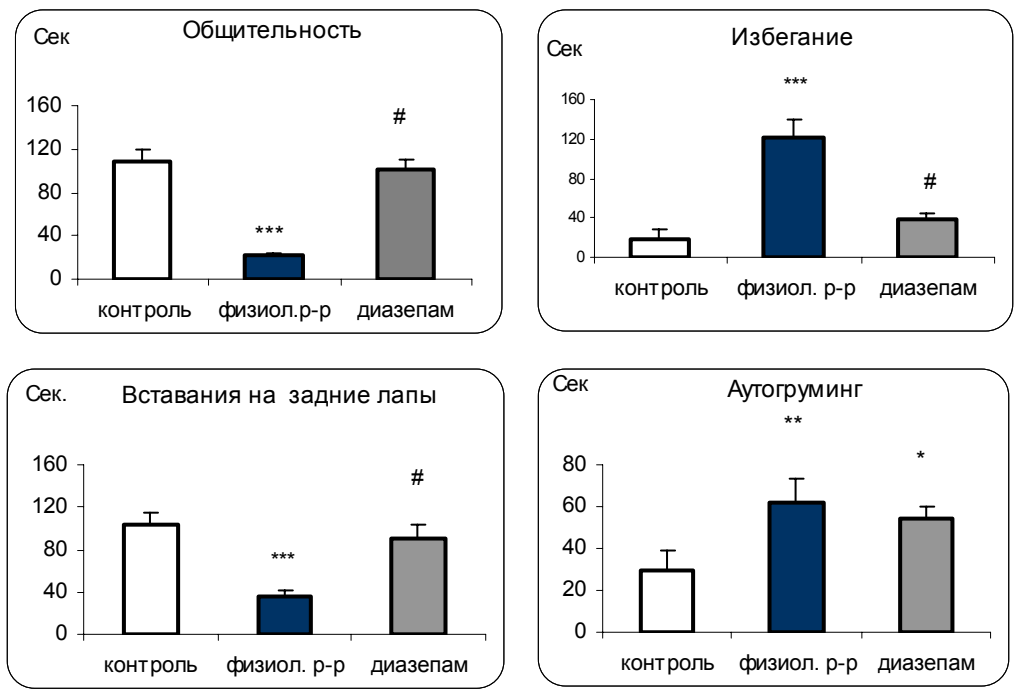

Рис. 2. Поведение самцов мышей экспериментальных групп (контроль, побежденные самцы после хронического введения физиологического раствора и диазепама) в тесте «социальные взаимодействия». Представлены данные по времени проявления соответствующих форм поведения.
\* - $p < 0.5$, \*\* - $p<0,01$, \*\*\* - $p<0,001$ – по сравнению с контролем; # - $p < 0.001$ – по сравнению с побежденными самцами, получавшими физиологический раствор

Figure. 2. Behavior of the controls (контроль) and defeated mice after chronic saline (физиол. р-р) and diazepam (диазепам) treatment) in social interaction test.
\* - $p < 0.5$, \*\* - $p<0,01$, \*\*\* - $p<0,001$ – vs controls; # - $p < 0.001$ – vs defeated mice, treated with saline

$p < 0,001$, соответственно). По сравнению с побежденными животными, которым вводили физиологический раствор, диазепам снизил число и время проявления избегания (для обоих показателей $p < 0.001$), и повысил параметры общительности (для обоих показателей *p* < 0.001), вставаний на задние лапы (*p* < 0.05 и *p* < 0.001, соответственно), но не аутогруминга (для обоих показателей, *p* > 0,05). Сравнение показателей поведения после введения диазепама с таковыми у контроля показало полное восстановление до нормы времени общительности (*p* > 0,05) и времени избегания (*p* > 0,05), числа и времени вставаний на задние лапы (для обоих показателей, *p* > 0,05). Не достигли уровня контроля после лечения диазепамом показатель числа демонстрации избегания партнера (*p* < 0.01) и общительности (*p* < 0.05). Данные по показателю числа форм поведения на рисунке не представлены.

## ОБСУЖДЕНИЕ

Как и ранее в наших экспериментах [6, 22, 23], повторный негативный опыт социальных взаимодействий привел к развитию у мышей высокого уровня тревожности, оцениваемого основными параметрами теста ПКЛ: число выходов и время нахождения в открытых рукавах и в центре были существенно меньше, а число выходов и время пребывания в закрытых рукавах у побежденных самцов были существенно больше, чем у контроля. Также были снижены число переходов из одного закрытого рукава в другой и параметров риска – выглядываний из закрытых рукавов и заглядываний под лабиринт. При этом, было обнаружено снижение уровня коммуникативности в тесте «перегородка» – время реагирования на знакомого и незнакомого партнера в соседнем отсеке у побежденных животных было существенно ниже, чем у контрольных особей. Снижение времени реагирования даже на незнакомого партнера в тесте «перегородка» может быть признаком избегания социальных контактов. Надо отметить, что снижение коммуникативности и повышение тревожности развивается у побежденных самцов с первых дней социальных взаимодействий.

Ранее было показано, что параметры поведения возле перегородки и параметры теста ПКЛ у интактных мышей являются коррелирующими показателями [24]. Это позволило нам утверждать, что поведение возле перегородки в реакции на другого партнера также может отражать и уровень тревожности. Корреляционный анализ показателей теста «социальные взаимодействия», с одной стороны, и показателей тестов «перегородка» и ПКЛ, с другой, не выявил достоверных взаимосвязей у контрольных животных по большинству параметров. Этот факт свидетельствует о том, что поведение в тесте «социальные взаимодействия» изначально не обусловлено тревожностью как таковой, а параметры теста отражают особенности поведения особей в новых условиях нейтральной клетки. Это и понятно, поскольку эмоциональный фон, по крайней мере, общительности (подходы к партнеру) и активного избегания и нежелания с ним взаимодействовать (замирание при подходе) противоположный – позитивный и негативный, соответственно. Достоверная корреляция была обнаружена в тесте «социальные взаимодействия» только для показателей времени и числа вставаний на задние лапы и времени демонстрации общительности, что позволяет предположить для этих форм поведения общую мотивационную составляющую – исследовательскую компоненту поведения, проявляющуюся и по отношению к незнакомому партнеру, и по отношению к новой ситуации.

После негативного социального опыта показатели теста «социальные взаимодействия» уже достоверно коррелируют с показателями тревожности в ПКЛ: чем больше такие животные заходят в закрытые рукава и меньше времени проводят в центре, тем больше они демонстрируют поведение избегания. Выраженность общительности и параметры тревожности в тесте ПКЛ, находятся в обратной зависимости. Таким образом, данные свидетельствуют о том, что негативный опыт социальных поражений вслед за развитием высокого уровня тревожности нарушает и социальные взаимодействия. Это прямо указывает на тревожность как фактор, приводящий к нарушению социабельности животных.

Хроническое введение широко используемого анксиолитика диазепама в течение двух недель снижало уровень тревожности, оцениваемый основными показателями теста ПКЛ, хотя и не всегда приближая его к таковому у контрольных животных. Однако нужно отметить, что дополнительные показатели – число выглядываний из закрытых рукавов лабиринта и число заглядываний под лабиринт, которые считаются показателями оценки риска, а также число переходов из одного закрытого рукава в другой практически не изменились под влиянием диазепама, сохраняясь на уровне таковых у побежденных животных, не прошедших лечения. Это свидетельствует о том, что эти формы поведения вряд ли включают в себя тревожность, как мотивационную компоненту поведения.

Под влиянием диазепама улучшается коммуникативное поведение в тесте «перегородка»: общее время пребывания возле перегородки стало практически на уровне контроля. После введения препарата число подходов к перегородке в реакции на незнакомого партнера снизилось по сравнению с контролем, однако это произошло за счет увеличения среднего времени пребывания возле перегородки за один подход. Это значит, что интерес к партнеру после диазепама стал даже несколько выше, чем у контрольных животных. Этот факт интересен сам по себе, позволяя предлагать диазепам в качестве препарата, восстанавливающего коммуникативность.

В тесте «социальные взаимодействия» тоже был выявлен позитивный эффект диазепама. Показатели общительности, исследовательской активности (вставания на задние лапы) и избегания партнера были равны аналогичным параметрам у контроля, возвращаясь после инъекций диазепама к норме. Таким образом, данные позволяют сделать вывод, что диазепам, снижая уровень тревоги, значительно улучшает показатели социального взаимодействия у побежденных самцов. В целом, можно сделать вывод, что фактором, вызывающим нарушение общительности, по крайней мере, в условиях нашей модели, является высокий уровень тревоги. При ее купировании восстанавливаются и социальные взаимодействия.

Поскольку нарушения в социабельности наблюдаются при аутизме, то возникал вопрос, в какой степени изменения в социальном поведении у мышей, вызванные влиянием негативного социального опыта, соответствуют тому, что наблюдается у людей. Согласно международной классификации болезней [4] диагностическими признаками аутизма является триада симптомов: недостаток социальных взаимодействий, нарушенная коммуникация и высокий уровень повторяющихся форм поведения (repetitive behavior). Эти основные симптомы могут сопровождаться ограниченностью интересов и изменением активности. По сути, в поведении побежденных самцов мышей просматриваются все симптомы аутистического спектра. Так, показано нарушение социальных взаимодействий. Побежденные животные демонстрируют, в основном, активные и пассивные формы избегания при подходе незнакомца и низкий уровень общительности, что свидетельствует не только о снижении коммуникативности, но и развитии неадекватного поведения даже по отношению к миролюбивому партнеру в ситуации, не несущей угрозы. Внимания заслуживает поведение аутогруминга, демонстрация которого у побежденных самцов выражена сильнее по сравнению с контрольными. Этот показатель мы рассматриваем, как правило, показателем смещенной активности у наших животных, помещенных в новые условия, но он может быть и маркером появления повторяющихся форм поведения, одного из симптомов аутизма. Все изменения в поведении, вызванные хроническим социальным стрессом, носят выраженный характер и не проходят даже после содержания животных в течение 14 дней без агонистических взаимодействий – различия между поведением у контрольных и побежденных животных все еще оставались значительными. Также у таких животных отмечали и сниженную исследовательскую активность: интерес к незнакомой особи и к новому нейтральному предмету в условиях домашней клетки [2] у них был существенно меньше, чем у контроля.

Таким образом, полученные данные дают основание полагать, что у животных под влиянием негативного социального стресса формируется состояние, при котором проявляются практически все основные симптомы аутистического спектра, появление которых вызвано высоким уровнем тревожности, развившейся в результате хронического социального стресса. Стоит отметить, что по разным сведениям тревожные расстройства находят у 11% - 84% детей с аутистическими симптомами [17, 30]. Сходное состояние у людей может быть классифицировано как социальный аутизм, то есть аутизм, вызванный негативным социальным опытом.

Хотелось бы обратить внимание на возможность использования нашего подхода и для изучения молекулярных механизмов аутизма. Наши предыдущие исследования показали, что повторный опыт социальных поражений приводит к увеличению экспрессии генов моноаминооксидазы и серотонинового транспортера в среднем мозге побежденных самцов [15]. Другие авторы показали изменение экспрессии многих других генов в разных структурах головного мозга у животных после однократного или повторного опыта социальных поражений [8, 9, 27]. При этом, сниженная коммуникативность [1, 8] и измененная экспрессия генов у животных могут длительно сохраняться после прекращения агонистических взаимодействий [8, 10].

Полагают, что расстройства аутистического спектра возникают под влиянием взаимодействия множества генов, либо при наличии редких мутаций [3] или эпигенетических изменений [28]. Предлагаются более сложные механизмы генетических влияний при аутизме [7, 16]. В то же время формируется мнение, что аутизм представляет собой сложное расстройство, вызванного многими причинами, часто действующими одновременно [18], причем, основы расстройства закладываются на ранних стадиях развития [5, 20], возникая, например, под влиянием пренатального стресса [19]. Полагают, что факторами предрасположенности к развитию аутизма могут выступать нарушения в диете, развитие некоторых болезней, токсические воздействия, инфекционные заболевания и множество других причин [обзор, 26]. Складывается впечатление, что необходимо уделять больше внимания внешним факторам среды как возможной причине развития аутизма, принимая во внимание не только генетические механизмы.

Можно предположить, что негативное воздействие, полученное в пренатальном и/или раннем постнатальном периоде развития индивида, может изменять экспрессию некоторых генов, приводя к появлению аутистической симптоматики. Изменение экспрессии генов может возникать и под влиянием среды в более поздний период. О возможности такого тотального воздействия свидетельствуют эксперименты по изучению молекулярных последствий хронического социального стресса. Изменения функционального состояния «специфических генов аутизма» или генов медиаторных систем, которые вовлечены в механизмы коммуникативности и/или ассоциированных форм поведения и длительно сохраняющие измененную экспрессию, могут создать видимость наследственно-обусловленного дефекта коммуникативности, который при выраженном проявлении может быть диагностирован как аутизм. Эти представления открывают новые перспективы в исследовании молекулярных механизмов аутизма и поиска способов его коррекции.

**Заключение.** Повторный опыт социальных поражений в агонистических взаимодействиях приводит к развитию выраженного уровня тревожности, снижению общительности, увеличению поведения избегания других особей, а также появлению повторяющих форм поведения у самцов мышей. Диазепам, вводимый побежденным животным в течение двух недель, существенно снизил тревожность и время избегания партнера, а также практически полностью восстановил уровень общительности и исследовательскую активность. Предполагается, что фактором, приводящим к выраженному нарушению социального поведения, является высокий уровень тревожности. Обсуждается сходство выявленных нарушений в социальном поведении у самцов мышей под влиянием социального стресса с симптомами аутистического спектра у людей.

## Литература